\documentclass[12pt]{article}

\usepackage{scicite}

\usepackage{psfig}

\newenvironment{sciabstract}{%
\begin{quote} \bf}
{\end{quote}}

\newcommand\degd{\ifmmode^{\circ}\!\!\!.\,\else$^{\circ}\!\!\!.\,$\fi}

\newcommand{\uv}{(u,v)}

\newcommand\aj{{\it Astron. J.}}%
\newcommand\araa{{\it Annual Rev. Astron. Astrophys. }}%
\newcommand\apj{{\it Astrophys. J.}}%
\newcommand\apjl{{\it Astrophys. J. Lett.}}%
\newcommand\apjs{{\it Astrophys. J.Supp.}}%
\newcommand\aap{{\it Astron. Astrophys.}}%
\newcommand\mnras{{\it Mon. Not. R. Astron. Soc.}}%
\newcommand\nat{{\it Nature}}%

\newcounter{lastnote}
\newenvironment{scilastnote}{%
\setcounter{lastnote}{\value{enumiv}}%
\addtocounter{lastnote}{+1}%
\begin{list}%
{\arabic{lastnote}.}
{\setlength{\leftmargin}{.22in}}
{\setlength{\labelsep}{.5em}}}
{\end{list}}

\title{Detection of the Intrinsic Size of Sagittarius A* through Closure Amplitude Imaging}

\author
{Geoffrey C. Bower,$^{1\ast}$Heino Falcke,$^{2}$ Robeson M. Herrnstein,$^{3}$\\
Jun-Hui Zhao,$^{4}$ W.M. Goss,$^{5}$, Donald C. Backer$^{1}$\\
\\
\normalsize{$^{1}$Astronomy Department \& Radio Astronomy Laboratory,}\\
\normalsize{University of California, Berkeley, CA 94720, USA}\\
\normalsize{$^{2}$Radio Observatory Westerbork,  ASTRON,}\\
\normalsize{P.O. Box 2 , 7990 AA Dwingeloo, The Netherlands}\\
\normalsize{$^{3}$Department of Astronomy, Columbia University,}\\ 
\normalsize{Mail Code 5246, 550 West 120th St., New York,NY 10027, USA}\\
\normalsize{$^{4}$Harvard-Smithsonian Center for Astrophysics,}\\ 
\normalsize{60 Garden Street, MS 78, Cambridge, MA 02138, USA}\\
\normalsize{$^{5}$National Radio Astronomy Observatory, 
Array Operations Center,}\\
\normalsize{P.O. Box O, Socorro, NM 87801, USA}\\
\\
\normalsize{$^\ast$To whom correspondence should be addressed; E-mail:  gbower@astro.berkeley.edu.}
}

\date{}

\begin{document}

\baselineskip24pt

\maketitle

\begin{sciabstract}
We have detected
the intrinsic size of Sagittarius A*, the Galactic Center radio source associated 
with a supermassive black hole,
 showing that the short-wavelength radio
emission arises from very near the event horizon of the black hole.
Radio observations with the Very Long Baseline Array show that the source
has a size of $24 \pm 2$ Schwarzschild
radii at 7 mm wavelength.  In one of
eight 7-mm epochs we also detect an increase in the intrinsic size of
$60^{+25}_{-17}$\%.  These observations place a lower limit to the
mass density of Sgr A* of
$1.4\times10^{4}$ solar masses per cubic astronomical unit.
\end{sciabstract}

Sagittarius A* (Sgr A*) is the compact, nonthermal radio source in the
Galactic Center associated with a compact mass of $4\times 10^6 M_\odot$
\cite{2001ARA&A..39..309M, 2003ApJ...586L.127G, 2003ApJ...596.1015S}.
It is the best established and
closest supermassive black hole candidate and serves as the prime test
case for the black hole paradigm.  Emission at radio, near-infrared, and X-ray  wavelengths traces processes in the environment of the event horizon
\cite{1998ApJ...499..731F,2003Natur.425..934G,2004ApJ...601L.159G,2001Natur.413...45B, 2002ApJ...571..843B,2003ApJ...588..331B}.

High resolution radio imaging of Sgr A* can ultimately distinguish between the many different models for the emission, accretion and outflow physics of the source as well as provide an important test of strong-field gravity \cite{2000ApJ...528L..13F}.  Sgr A* has been a target of such observations for the past 30 years \cite{2003inprepgoss}.  Its intrinsic size and structure have remained obscured, however, because radio waves from Sgr A* are scattered by turbulent interstellar plasma along the line of sight 
\cite{1998ApJ...505..715L}.  
The scatter-broadened image of Sgr A* is an ellipse with the major axis 
oriented  almost exactly East-West and a quadratic size-wavelength relation.

The turbulent plasma is parametrized with a power-law of turbulent energy density as a function of length scale
with outer and inner scales that correspond to the scale on which turbulence is generated and damped, respectively.  Scattering theory predicts that the scatter-broadened image will be a Gaussian when the inner length scale of the turbulent medium is larger than the longest baseline of the observing interferometer \cite{1989MNRAS.238..963N}.  Additionally, the scatter-broadened image size will scale quadratically as a function of wavelength.  In the case of Sgr A*, the longest interferometer baseline used in our analysis $b_{max}\sim 2000$ km  corresponds to a length scale in the scattering medium $D_{scattering}/D_{source}\times b_{max}\sim 25$ km, where $D_{source}=8$ kpc is the distance from Sgr A* to the Earth and $D_{scattering}=100$ pc is the distance from Sgr A* to the scattering screen
\cite{1998ApJ...505..715L}.    This scale is much less than the predicted and measured values of the inner scale,
which fall in the range $10^2$ to $10^{5.5}$ km \cite{1994MNRAS.269...67W, 2001ApJS..133..395D}.  The amplitude of turbulence in the Galactic Center scattering screen is $\sim 2-3$ orders of magnitude greater than what is seen in the next most powerful scattering region, NGC 6334B\cite{1998ApJ...493..666T}, however, suggesting that the Galactic Center case may be atypical.  

The presence of strong scattering has pushed observations to shorter and
shorter wavelengths where scattering effects decrease and intrinsic source 
structure may dominate, creating a deviation from the
measured size-wavelength law.  On the basis of extensive
observations with the National Radio Astronomy Observatory's
Very Long Baseline Array (VLBA), L98 measure the index of the size-wavelength power-law to be $\alpha=1.99 \pm 0.03$\cite{1998ApJ...508L..61L}.  L98 also claim a deviation from the scattering
law in the minor axis at 7mm wavelength (43 GHz), implying an intrinsic size of 72 Schwarzschild radii
($R_s$)\cite{size}.

Unfortunately, precise measurements of the size of Sgr
A* are seriously
hampered by calibration uncertainties related to
the variable antenna gain and atmospheric opacity at the low antenna elevations
necessary to observe Sgr A* from the northern hemisphere.  
Closure amplitudes have been used to
constrain the size of Sgr A* with VLBI observations at 3.4 mm \cite{2001AJ....121.2610D}.
The closure amplitude does not
rely on calibration transfer from another source as traditional imaging methods
do and is independent of
all station-dependent amplitude errors.    This method does not, however, eliminate
baseline-dependent errors such as variable decorrelation (which also
influence conventional calibration and imaging techniques). The closure amplitude is conceptually related to the closure phase, a more
well-known quantity which is also independent of station-based gain errors.
 The principle
drawback of closure amplitude analysis for simple source structures is
the reduction in the number of degrees of freedom relative to a
calibrated data set.  The number of independent data points for a
7-station VLBA experiment is reduced by a factor 14/21.  Additionally,
the closure amplitude method can not determine the absolute flux
density for the source.  These shortcomings are more than offset
by the confidence that the result gives through its accurate handling
of amplitude calibration errors.

We describe here the analysis of new and archival VLBA data through
closure amplitude and closure phase quantities.  We analyze 3 new
experiments including data at 1.3 cm, 6 new experiments including data
at 0.69 cm, as well as 10
experiments from the VLBA archive including data at 6, 3.6, 2.0, 1.3, 0.77, 0.69 and 0.67 cm wavelength.

\section*{Observations and Initial Data Reduction}

Six new observations were made with the VLBA as part of our Very Large Array
flux density monitoring program \cite{2004inprepherrnstein}.
Three
observations were made in each of two separate epochs in July/August 2001 and April/May 2002
(STable~1).
In the first epoch, observations at 1.3 cm and 0.69 cm were interleaved over 5 hours.
In the second epoch, observations were obtained only
at 0.69 cm in order to maximize the signal to noise ratio (SNR) of the final result.
All observations were dual circular polarization with 256 Mbits/sec recording rate.

We also analyzed a number of experiments from the VLBA archive over the wavelength range of 6.0 cm to 0.67 cm (STable~1).
The experiments BS055 A, B and C were
those analyzed by L98.  The experiment BB113 was previously analyzed 
\cite{2001ApJ...558..127B}.

Initial data analysis was conducted with the NRAO Astronomical Imaging Processing System
\cite{2003iha..book..109G}.
Standard fringe-fitting techniques were employed to
remove atmospheric and instrumental delays from the data (SOM text).
High SNR fringes were detected for most stations on the compact source NRAO 530
(J1733-1302), indicating the overall quality of the data.  Due to the relatively
larger size of Sgr A*, fringes were obtained for a subset of 5 to 8 stations
(STable~1).

Data were then averaged over wavelength and time for each experiment.
The quality of the final result is dependent upon the visibility averaging time.
The longer the averaging time, the higher the SNR of the closure amplitude
calculation \cite{1995AJ....109.1391R}.
On the other hand, as the averaging time approaches the phase
decorrelation time, the closure amplitudes
cease to be accurate.  It is not necessary, however, to determine the best averaging time 
precisely, since neither of these effects is a strong function of time
\cite{1995AJ....109.1391R}.  The results that we give are for an averaging
time of 30 seconds, but we find that for averaging times of 15 to 120 seconds
the estimated intrinsic size of Sgr A* does not differ by more than 10\%
(SOM text).
No amplitude calibration was applied at
any stage.  The averaged data were then written to text files for analysis by our
own analysis programs, external to AIPS.

\section*{Closure Amplitude and Closure Phase Analysis of a Single Gaussian}

We
form the closure amplitude from the measured visibilities and  average the closure amplitudes over time.
Closure amplitudes were averaged over scans, which were 5 to 15 minutes in duration.
The code uses the scatter in the closure amplitudes before averaging to determine the
error in the closure amplitude.  Only independent closure amplitudes were formed
\cite{2001isra.book.....T}.

We selected visibility data only with station elevations $>10^{\circ}$ to reduce
sensitivity to phase decorrelation, which is more significant at low elevations.
We also excluded data at $\uv$ distances greater than 25 $M\lambda$ at 6.0 cm, 50 $M\lambda$ at 3.6 cm, 150 $M\lambda$ at 2.0 cm and 1.3 cm, and 250 $M\lambda$
at 0.69 cm.  These sizes are comparable to the expected size of Sgr A* at each wavelength.
Visibility amplitudes beyond the cutoff were indistinguishable by inspection from noise.
This $\uv$-distance limit reduced sensitivity
to the noise bias or station-dependent differences in the noise bias.
Results were not strongly dependent on the value of this cutoff.

Model visibilities for each baseline and time datum were computed for an elliptical source
of a given flux density $S_0$, major axis size $x$, minor axis size $y$ and position angle
$\phi$.  In addition,
a noise bias was added in quadrature to each model visibility.  Our model visibility amplitude
(squared) on baseline $ij$ is then
\begin{equation}
A_{ij}^2 = S_0^2
           e^{ - D_0
           ( (u_{ij}^{\prime}  x)^{\beta -2} + (v_{ij}^{\prime}  y)^{\beta -2} ) }
           + N_{ij}^2,
\label{eqn:gaussian}
\end{equation}
where $D_0= 2 ({\pi \over {2 \sqrt{ \log{ 2}}}} )^2$,
$N_{ij}$ is the noise bias,
and $u_{ij}^\prime$ and $v_{ij}^\prime$ are
baseline lengths
in units of wavelength in a coordinate system rotated to match the position angle $\phi$.
Model closure amplitudes were
then formed from these model visibilities.  We determine the best-fit parameters using
a non-linear fitting method that minimizes $\chi^2$ between the model and measured
closure amplitudes (SFig.~1,STable~2).
We find the reduced $\chi^2$ for the amplitudes $\chi^2_A \approx 1$ for all experiments.

In the case of an image produced by interstellar electron scattering on baselines longer than the inner scale of turbulence,
$\beta$ is the power-law index of electron density fluctuations
\cite{1989MNRAS.238..963N}. The parameter $\beta$ is related
to the exponent $\alpha$ of the scattering law
(size\ $\propto\lambda^\alpha$) as $\beta=\alpha+2$, allowing an independent check of
the $\lambda^2$ law
\cite{1989MNRAS.238..963N,1994MNRAS.269...67W,2001ApJS..133..395D}.
For the case of the Galactic
Center scattering we expect $\beta=4$, in which case
Equation~\ref{eqn:gaussian} is a Gaussian function
and $x$ and $y$ are the FWHM in the two axes.
Allowing $\beta$ to be unconstrained in our fits, 
we find $\beta=4.00 \pm 0.03$, which is
consistent with the expectation of
scattering theory (SFig.~3).  All remaining analysis is conducted with the assumption
that $\beta=4$.

The introduction of the noise bias to the model changes our calculation from
a pure closure amplitude to a noise-biased closure amplitude.
We found that our results did not
require that we consider  the noise bias as dependent on station or time
(SOM text).
Thus, we chose $N_{ij}(t) = N_0$
because it is simpler computationally and has a smaller number of independent
parameters.

Errors in the model parameters were determined
by calculating $\chi^2$ for a grid of models surrounding the solution and fitting constant $\chi^2$
surfaces (SFig.~2).  
Monte Carlo simulations find confidence intervals that are smaller by a factor of two than
determined from the $\chi^2$ analysis, suggesting that the dominant sources of error
are  baseline-based
errors such as phase decorrelation, which were not included in the Monte Carlo simulations
(SOM text).

Closure phases were formed, averaged and analyzed in a manner similar
to the closure amplitudes.  We tested the closure phases against the hypothesis
that they are all zero.  This hypothesis is the case for a single elliptical
Gaussian and other axisymmetric structures with sufficiently smooth brightness
distributions.  An axisymmetric disk  is a notable exception to this hypothesis since
it induces ringing in the transform plane.
The reduced $\chi^2$ for this hypothesis
$\chi^2_\phi\approx 1$
for all experiments (STable~2), indicating no preference for multiple components,
non-axisymmetric structure or disk-like structure.

Although the solutions for a single Gaussian component are sufficiently accurate, we
did search the parameter space for two component models.  To do this, we performed
a minimization of $\chi^2$ with respect to closure amplitude and closure phase
jointly.
The reduced $\chi^2$ for these models was roughly equal to the values for the single Gaussian
component despite the addition of several degrees of freedom.  We also calculated upper limits
to the flux densities of secondary components that are in the range 2-10\%, typically
(SFig.~4,STable~2).  The absence of any improvement
indicates that a single Gaussian component is sufficient and the simplest model
of the data.  This absence is particularly significant for the cases where $\chi^2>1$ and suggests,
as noted before, that the results are dominated by closure errors rather than improperly
modeled structure.

\section*{Scattering Law and Intrinsic Size}

We determined the size of the major and minor axes of Sgr A* for each experiment
(Fig.~1 and 2, STable~2).  The major axis is oriented almost exactly
East-West.
The major axis size is
measured much more accurately than the minor axis size because
of the poorer North-South resolution of the array.  All major axis measurements at 1.3 and 0.69 cm are larger than the scattering
size determined by L98\cite{L98} and the new scattering size that we determine below, although the difference is statistically significant in only one epoch at 0.69 cm.
Minor axis measurements are distributed about the scattering
result and no one differs significantly from the expected result.

The L98 scattering law is adequate for the minor axis measurements as
a function of wavelength
(Fig.~3).
All the measured minor axis sizes agree with the scattering
law to better than $3\sigma$.
The data are also consistent with a constant position angle
of $78.0^{+0.8}_{-1.0}$ deg with $\chi_{\nu}^2=2.2$
for $\nu=6$ degrees of freedom.

We determine fits to the major and
minor axis sizes as a function of wavelength using subsets of the data with a minimum
wavelength $\lambda_{min}$ of 2.0 cm, 1.3 cm, 0.6 cm and 0.3 cm (STable~3).  The
last fit includes the 3.4 mm circular Gaussian fits of
for the major axis only \cite{2001AJ....121.2610D}.
There are two fits for each
subset, allowing $\alpha$ to vary and fixing $\alpha=2$.  $\chi_{\nu}^2$ is less
than 3 for the minor axis case with $\lambda \geq 0.6$ cm, confirming that the
solution is adequate for $\alpha=2$.

The major axis data, however, are discrepant from the
L98 and the new scattering law (Fig.~3).  All of the
7mm results fall above the L98 scattering law.  Two of these points
are significantly different at greater than 3$\sigma$.  The L98 scattering
law predicts a size of 690 $\mu$arcsec at 0.69 cm, which is $\sim
7\sigma$ from the measured size ($712^{+4}_{-3}$ $\mu$as)
and smaller than any of the measured
sizes (Fig.~2).  An attempt to fit a scattering law
with $\alpha_{\rm major}=2$ to all data with $\lambda > 0.6$ cm gives
$\chi_{\nu}^2=24$ for 6 degrees of freedom, demonstrating that the
hypothesis can be strongly rejected.  In fact, the 1.35 cm major axis
size is also discrepant with the best-fit $\alpha_{\rm major}=2$
scattering law, giving $\chi_{\nu}^2=5.6$ for 3 degrees of freedom.

We consider two alternative models for our resuls:  case A, the scattering power-law exponent $\alpha_{\rm major}$ is not exactly 2;
or, case B, intrinsic structure in Sgr A* is distorting
the size-wavelength relation at short wavelengths.

For case A, we find 
adequate solutions for all data at wavelengths $\geq 0.3$ cm with
$\alpha_{\rm major}=1.96 \pm 0.01$.  The result is clearly discrepant with
scattering theory which requires $\beta=4$ and marginally discrepant with our determination
of the scattering theory
parameter $\beta=4.00 \pm 0.03$ (SFig.~3),
since scattering theory predicts that $\alpha=\beta-2$.

For case B, we determine a new  scattering law from observations
with $\lambda \geq 2.0$ cm and $\alpha_{\rm major}=2$.  This solution has a scale
parameter $\sigma_{\rm major}^{\rm 1cm}$
that is even less than that of L98, increasing the discrepancy at short wavelengths.
Removing this new scattering law in quadrature
gives an intrinsic size of $0.7 \pm 0.1$ mas at 1.35
cm, $0.24 \pm 0.01$ mas at 0.69 cm and $0.06 \pm 0.05$ mas at 0.35 cm
(Table~1).
On the basis of the disagreement between $\beta$ and $\alpha$, we reject
case A and claim that we have determined the size of intrinsic
structure in Sgr A* at 1.35 and 0.69 cm.

The two cases predict substantially different sizes at 20 cm.
For the major axis case A predicts $541 \pm 2$ mas while case B predicts $595 \pm 3$.
The 20.7 cm ($\nu=1450$ MHz)
major axis size $624 \pm 6$ mas measured with the VLA A-array 
\cite{1994ApJ...434L..63Y}
is discrepant with both of these cases, although more strongly with case A.
These measurements are particularly difficult since the source is only partially
resolved in the A-array:  the synthesized beam is about $2.6 \times 0.9$ arcsec oriented
North-South.  Additionally, extended structure in the Galactic Center makes estimation
of the size strongly dependent on the estimate of the zero-baseline flux density.

We attempted to verify the 20 cm size
with analysis of three VLA A-array observations
at 21.6 cm obtained originally for polarimetry
\cite{2002ApJ...571..843B}.  Results for each of the three experiments were similar and
dominated by systematic errors that make an estimate of the intrinsic
size difficult.  We were unsuccessful at analyzing these
experiments with our closure amplitude technique, possibly due to the poor resolution of Sgr A* and inability of our code to handle the large number of stations.  In any
case, the reliability of amplitude calibration of the VLA at 20 cm
reduces the need for closure amplitude analysis.  We imaged
all baselines and measured the total flux density of Sgr A* by fitting a
two-dimensional Gaussian to the central 3$^{\prime\prime}$.  For all epochs, we find an error in the total flux density of 10 mJy.  We determined the size by fitting in the $\uv$ plane with the total flux density fixed and with a minimum cutoff in $\uv$ distance.  For values of the total flux density that range from $-1\sigma$ to $+1\sigma$ and for a minimum $\uv$ distance from 
$20$ to 120 $k\lambda$, we find that the major axis size varies systematically
from 580 to 693 mas.  The
minor axis is very poorly constrained.  We estimate the size from the
mean of these results as $640 \pm 40$ mas.  We consider  this to be a more reasonable estimate of the error in the size of Sgr A* than previously given.  This size is consistent at
$<1\sigma$ with case B and $\sim 1.5\sigma$ with case A, favoring slightly
detection of the intrinsic size.

Although all minor axis data are adequately fit with $\alpha_{\rm minor}=2$,
we can check the consistency of our results
by estimating intrinsic sizes for this axis in the same way.
The minor axis sizes show the same trend
as the major axis sizes:  smaller than the L98 scattering law at long wavelengths
and larger than the L98 scattering law at short wavelengths (Fig.~3).
Using the solution for $\alpha_{\rm minor}=2$ and $\lambda \geq 2.0$ cm,
we estimate intrinsic sizes of $1.1 \pm 0.3$ mas at 1.35 cm
and $0.26 \pm 0.06$ mas at 0.69 cm.  These are comparable to the sizes
determined for the major axis.  For the case of unconstrained power-law index fit to all data, we find $\alpha_{\rm minor}=1.85^{+0.06}_{-0.06}$, marginally
consistent with no intrinsic source.

\section*{Changes In the Source Size with Time}

At 0.69 cm, the only measurement deviating significantly from the mean result is in the major axis for BB130B.  The BB130B result is $770^{+30}_{-18}\ \mu$as while
the mean result is $712^{+4}_{-3} \ \mu$as giving a difference
of $58^{+30}_{-19} \ \mu$as.  We note that the greatest deviation in
the 0.69 cm position angle also occurs for BB130B, although the
difference is significant only at the $2\sigma$ level. Any such
deviation would indicate a non-symmetric expansion or a non-symmetric
intrinsic source size.  We can estimate the change in the size of the intrinsic source between
BB130B and the mean size by subtracting in quadrature the case B
scattering size from each.  As stated above, the mean result implies
an intrinsic size of $0.24 \pm 0.01$ mas.  The intrinsic size implied
by the BB130B result is $0.38^{+0.06}_{-0.04}$ mas.  Thus, the growth
in major axis size is $0.14^{+0.06}_{-0.04}$ mas in the N-S direction.
We cannot associate this change in structure with a flux density change.
This maximum in the size comes $\sim 10$ days before detection of an
outburst at 0.69 cm with the VLA \cite{2004inprepherrnstein}.
The following epoch, BB130C, occurs only two days before this outburst
but shows no deviation from the mean size, although the size is
particularly poorly determined in this case.

\section*{The Interstellar Scattering Screen}

The image of a scattered source is created by turbulent plasma along the line of sight.  The minimum time scale for the scattered image to change is the refractive time scale, the time in which the relative motions of the observer, turbulent plasma and background source lead to the background source being viewed through a completely different region of the interstellar plasma.  The refractive time scale for Sgr A* is $\sim 0.5 \lambda^2
{\rm\ y\ cm^{-2}}$ given a relative velocity of 100 ${\rm\ km\ s^{-1}}$
\cite{1989MNRAS.238..963N}.
At our longest wavelength for VLBA observations, 6 cm,
then the time scale is 20 y.  At our shortest wavelength of 7 mm, the time scale
for refractive changes is 3 months.  Our observations are distributed over a
much larger time frame than three months, implying that the mean result
may be affected by refractive changes.

Two subsets of the archival data
have much smaller span, however.  The BS055 experiments cover 6.0 to 0.69 cm in
1 week and the BL070 and BB113 experiments cover 6.0 cm to 0.67 cm in 3 months.
These data sets include all of the 2.0 cm and longer wavelength data.
If we compare the 0.7 cm size, we see
that it is larger in these quasi-simultaneous
experiments than in the mean of all experiments and
also larger than the expectation of the new scattering law.  
We find 0.69 cm major axis sizes of $728^{+16}_{-11}\ \mu$as and $713^{+12}_{-9}\ \mu$as for
BS055C and BL070B, respectively,  both larger than the mean size
of $712^{+4}_{-3}\ \mu$as (STable~2).  We conclude
that if refractive
effects are altering the short wavelength results,  then
their effect is to reduce the deviation from the scattering
law, not enhance it.

\section*{Discussion}

Our results allow us to probe the mechanisms responsible for accretion,
outflow and emission in the vicinity of the black hole.
We can compare the measured 7mm intrinsic major axis size of
$24 R_s$  and its dependence on wavelength with expected values (Fig. 4).
The intrinsic size of the major axis decreases with wavelength and
is best-fit with a power-law
as a function of wavelength with index $\alpha_{\rm intrinsic} =1.6 \pm 0.2$.
We find for the minor axis a similar value $\alpha_{\rm intrinsic} =2.1 \pm 0.5$.
Assuming that the source is circularly symmetric and using the
mean flux density of 1.0 Jy at 7mm \cite{2004inprepherrnstein},
we compute a brightness temperature
$T_b=1.2 \times 10^{10} \times \left( \lambda \over {\rm 0.7 cm} \right)^{-1.2}$ K.  
This result is a lower limit, because the source
may be smaller in the minor axis.  A brightness temperature in excess of $10^{10}$ K
is a strong indication that synchrotron radiation is the dominant emission mechanism 
at work.

The wavelength-dependent size of Sgr A* now unambiguously
shows that the source is stratified due to optical depth effects.  We
rule out models in which the emission originates
from one or two zones with  simple  mono-energetic electron distributions
\cite{1997A&A...328...95B}. These models predict
a size which is constant with wavelength and is larger than our measured size.

The results are well-fit by a multi-zone or inhomogeneous model, in which the size is equal to the radius at which the optical depth is equal to unity
\cite{1979ApJ...232...34B}.   In a jet model, declining magnetic field strength, electron density and electron energy density contribute to a size that becomes smaller with wavelength.  A detailed jet model for Sgr A* predicts an intrinsic size of 0.25 mas at 0.69 cm and 0.6 mas at
1.3 cm (Fig. 4) \cite{2000A&A...362..113F}.  Exact values and wavelength
dependence are a function of  a
number of parameters including the relative contributions of the extended
jet and the compact nozzle component of the jet.
The jet model also predicts that the source should be elongated with an axial
ratio of 4:1.  The apparent measured symmetry in the deconvolved sizes in each axis, however, does not imply
that the intrinsic source is symmetric.  For example, an elongated intrinsic source that is oriented at 45 degrees to the scattering axis will produce equal deconvolved
sizes  in each axis.  Modeling of the closure amplitudes with a complete source and scattering model is necessary to determine the elongation for the most general case.

The thermal, high accretion rate  models such as Bondi-Hoyle accretion
\cite{1994ApJ...426..577M} and advection dominated accretion flows
\cite{1998ApJ...492..554N} require $T_e \sim 10^9$ K, which
overpredicts the size in each axis by a factor of 3.  This disagreement confirms the elimination of
these models on the basis of the polarization properties of Sgr A*
\cite{2003ApJ...588..331B}.  On the other hand,
the radiatively inefficient accretion flow (RIAF) model
\cite{2003ApJ...598..301Y}
has a lower accretion rate and higher $T_e$, compatible with the
polarization and with this measurement.  The RIAF model also predicts an inhomogeneous electron distribution consistent with a size that reduces with decreasing wavelength.  Both the RIAF model and the jet model are similar in the electron energy distribution and magnetic field distribution required to produce the observed flux density within the observed size.  These models differ principally in the relative contribution of thermal electrons to emission in the submillimeter region of the spectrum.

Extrapolating our size-wavelength relation to longer wavelengths, we estimate a size at 2 cm of 130 $R_s$ with a characteristic light travel time of 85 minutes.  This is comparable to the shortest
time scale for radio variability detected, 2 hours, during which the 2.0 cm
radio flux density changed by 20\% \cite{2002ApJ...571..843B}.  The smooth nature of the spectrum from 90 cm to 7 mm, suggests that our size-wavelength relation
holds over that entire range
\cite{2004ApJ...601L..51N}.  

Our relation implies a
size $<2 R_s$ at 1.3mm, comparable to the size of the event horizon.  The decrease of the source size with
wavelength cannot continue much farther due to the finite size of the
central object itself.  In the millimeter and submillimeter, however, the spectral index rises \cite{2003ApJ...586L..29Z}, indicating that there may be a break in the size-wavelength relation.  Ultimately, the size of the event horizon can be viewed as setting a limit on the wavelength of the peak emission.  The strong break in the spectrum between the submillimeter and the NIR may correspond to the wavelength at which the source size becomes comparable to  the event horizon.  Even with a weaker dependence of size on wavelength, the light travel time scale at millimeter wavelengths is a few minutes, comparable to
the shortest time scale observed at X-ray and NIR wavelengths.  This coincidence suggests
that the bright flares observed in at higher energies
\cite{2001Natur.413...45B,2003Natur.425..934G,2004ApJ...601L.159G}
are related to the submillimeter part of the spectrum and come
from the vicinity of the black hole.  The proximity of the millimeter
emission indicates 
that emission at this and shorter wavelengths will be subject to
strong light bending effects, providing a unique probe of strong-field
general relativity \cite{2000ApJ...528L..13F,2003MNRAS.342.1280B}.

The size-wavelength relation also implies that the black hole
mass must be contained within only a few Schwarzschild radii.
Radio proper motion measurements require that 
Sgr A* must contain a significant fraction 
if not all of the compact dark
mass found in the Galactic Center 
\cite{1999ApJ...524..805B,1999ApJ...524..816R,Reidetal2003}.
Using conservatively only our 7 mm size and the lower limit of the Sgr A*
mass of $4\times10^5M_\odot$,
we find that the mass density in Sgr A* has to be strictly
above $\rho_\bullet>1.4\times10^{4}M_\odot$AU$^{-3}$.  The dynamical
lifetime of a cluster of objects with that density would be less than
1000 years, making Sgr A* the most convincing existing case for a massive black
hole\cite{1998ApJ...494L.181M}.


\begin{thebibliography}{10}
\bibitem{L98}
L98 determined a scattering law $\sigma_{\rm axis} = \sigma_{\rm axis}^{\rm
  1cm} \lambda^{\alpha_{\rm axis}}$, where axis is either major or minor,
  $\lambda$ is given in cm, $\sigma_{\rm major}^{\rm 1c m}=1.43 \pm 0.02$ mas,
  and $\sigma_{\rm minor}^{\rm 1cm}=0.76 \pm 0.05$ mas. $\alpha$ is the index
  of the scattering law and it is assumed to be 2 for the strong scattering
  case. L98 found $\alpha_{\rm major}=1.99 \pm 0.03$. The major axis is
  oriented almost purely East-W est at a position angle of 80$^\circ$ East of
  North.
\end{thebibliography}

\begin{scilastnote}
\item The National Radio Astronomy Observatory is a facility of the National Science Foundation
operated under cooperative agreement by Associated Universities, Inc.
\end{scilastnote}

\clearpage

\begin{tabular}{rrrr}
\multicolumn{4}{c}{{\bf Table 1.} Intrinsic Size of the Major Axis of Sgr A*} \\
\hline
\multicolumn{1}{c}{Wavelength} & \multicolumn{1}{c}{Measured Size} & \multicolumn{1}{c}{Scattering Size} & \multicolumn{1}{c}{Intrinsic Size}  \\
(cm)       &   ($\mu$as)       &   ($\mu$as)         &    ($R_s$) \\
\hline
1.35	   &   $2635^{+37}_{-24}$ & $2533^{+20}_{-20}$ & $72^{+15}_{-11}$ \\
0.69 	   &   $712^{+4}_{-3}$    & $669^{+5}_{-5}$ & $24^{+2}_{-2}$ \\
0.35     &     $180^{+20}_{-20}$  & $173^{2}_{-2}$   & $6^{+5}_{-5}$ \\
\hline
\end{tabular}

\clearpage

\newpage
\begin{figure}[t]
\mbox{\psfig{figure=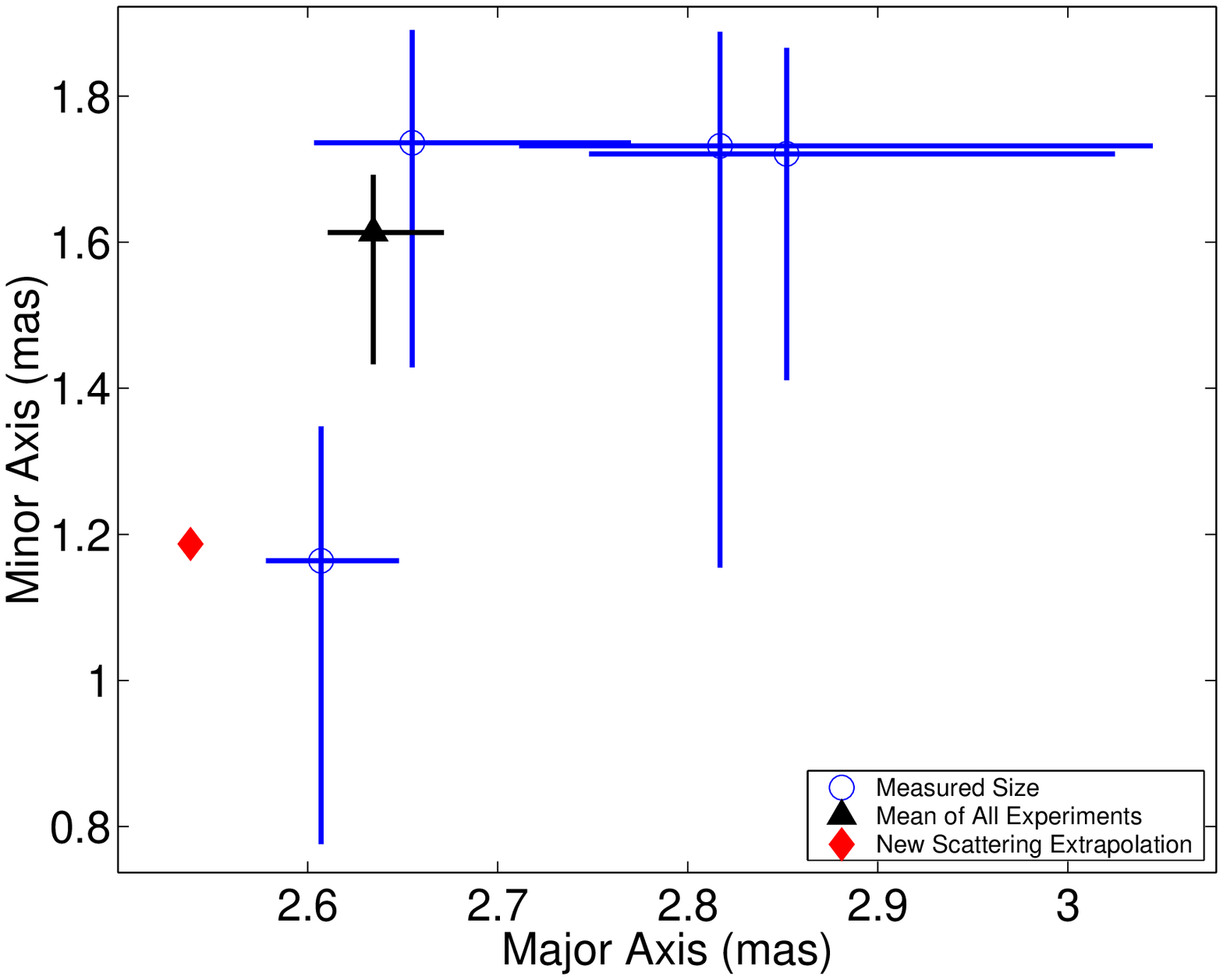}}
\end{figure}
\noindent {\bf Fig. 1.}{Sizes from closure amplitude analysis of 3 new and 1 archival 1.3 cm VLBA experiments (open circles).  The mean size (triangle)
is significantly larger than the new scattering size (black diamond), which is a fit to all
data at $\lambda \geq 2$ with $\alpha=2$.
\label{fig:results_k}
}
\newpage
\begin{figure}[t]
\mbox{\psfig{figure=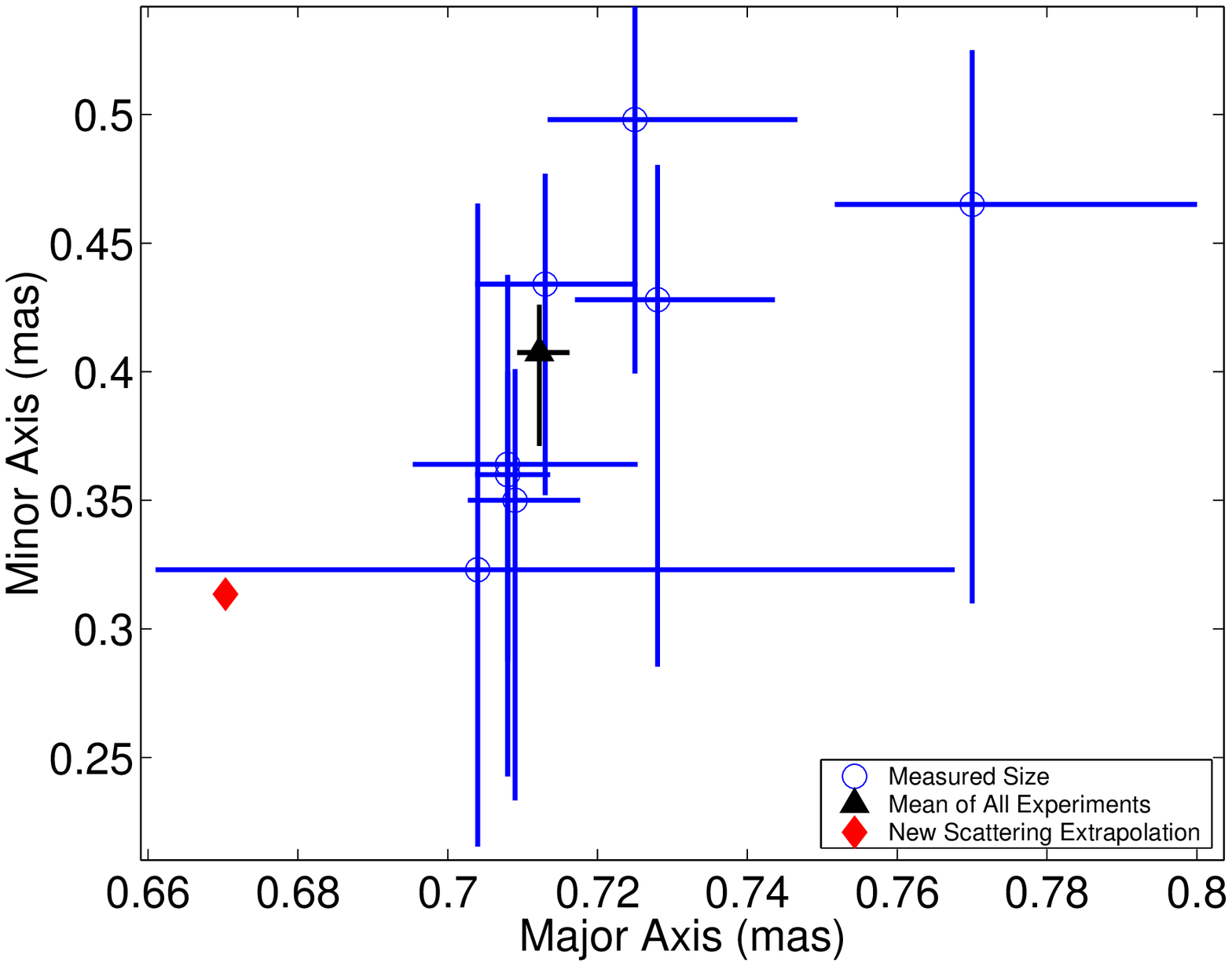}}
\end{figure}
\noindent {\bf Fig. 2.}{Sizes from closure amplitude analysis of 6 new and 2 archival 0.69 cm VLBA experiments (open circles).  
The mean size (triangle) is significantly larger than the new scattering size (black diamond).
\label{fig:results}
}
\newpage
\begin{figure}[t]
\mbox{\psfig{figure=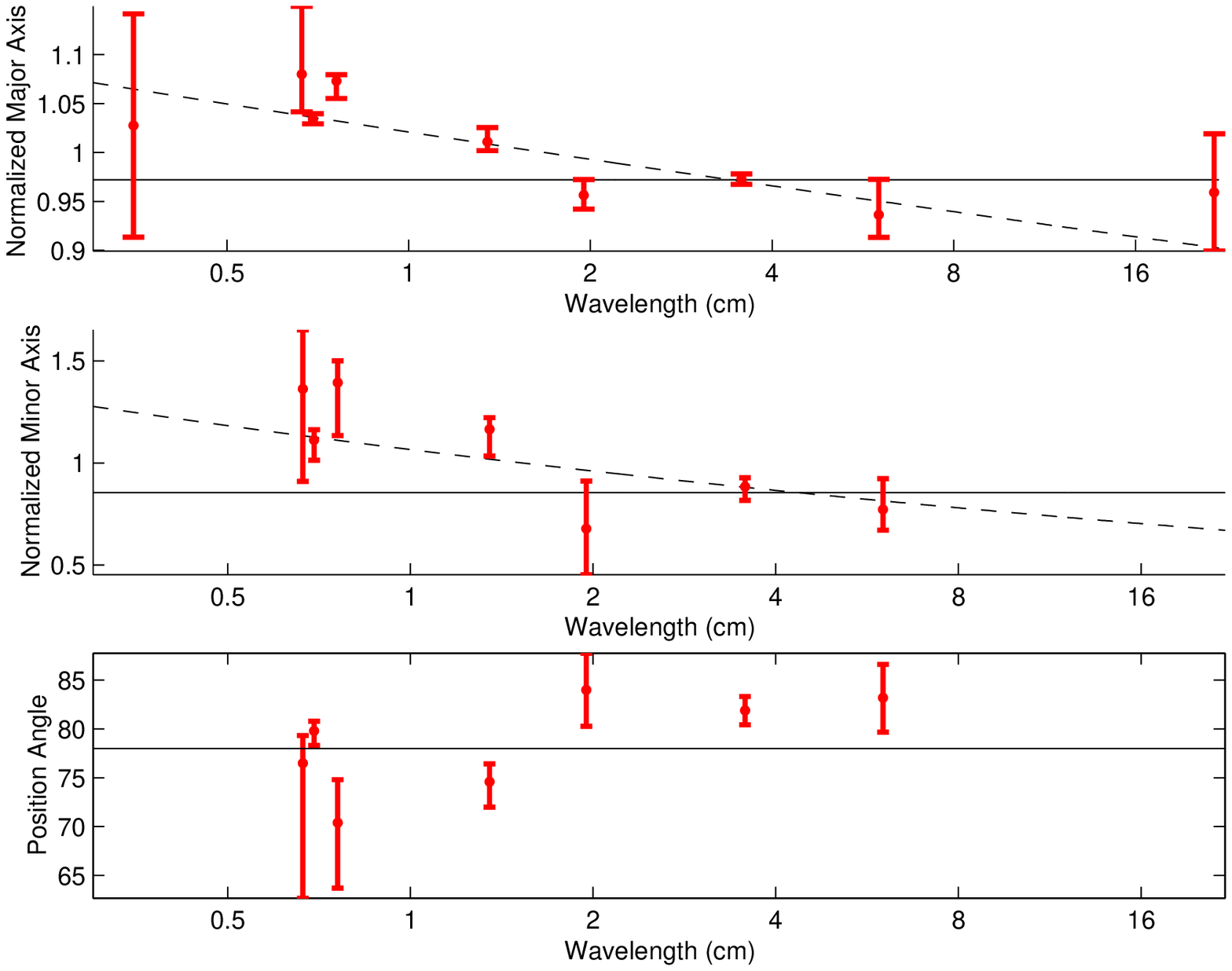}}
\end{figure}
\noindent {\bf Fig. 3.}{Major axis size, minor axis size and position angle
as a function of wavelength normalized to the L98 scattering
size.  All results  are determined by the closure amplitude technique
except the 21.6 cm result, which is determined from conventional fitting.
We also include a 3.5 mm measurement in the major axis
\cite{2001AJ....121.2610D }.  In the upper two panels, solid lines show the best-fit
$\lambda^2$ scattering model for $\lambda \geq 2$ cm in the major
and minor axis plots.  The dotted lines show the best-fit law
with $\alpha$ unconstrained for $\lambda \geq 0.6$ cm.
The line in the lower panel is our best fit value of 78 degrees for the position angle.
\label{fig:angsize}}
\newpage
\begin{figure}[t]
\mbox{\psfig{figure=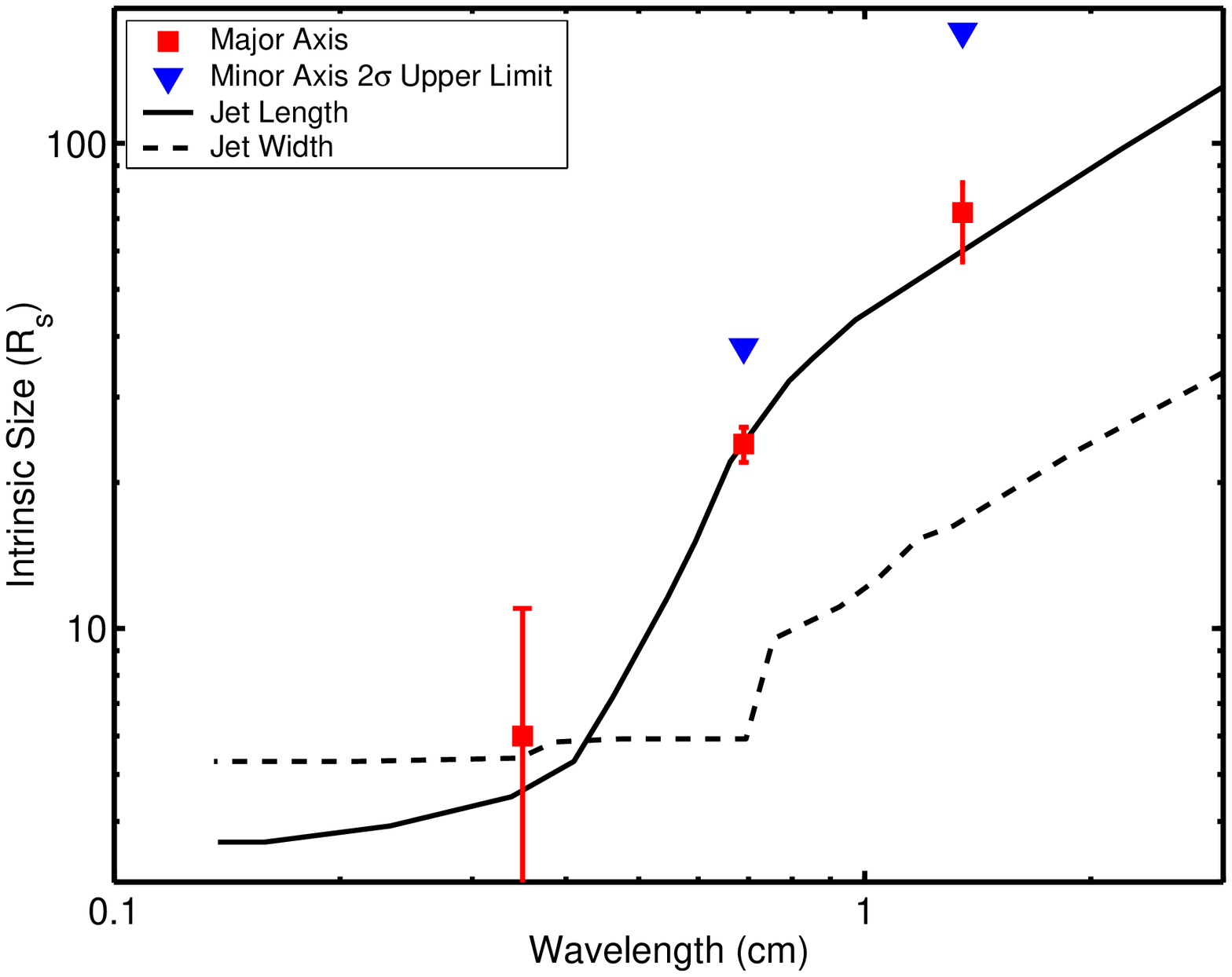}}
\end{figure}
\noindent {\bf Fig. 4.}{The intrinsic size of Sgr A* as a function of wavelength.
We plot the best-fit size in the major axis and $2\sigma$ upper limits to the size of the minor axis.
We also plot one set of predictions for the jet length and jet width 
\cite{2000A&A...362..113F}.
\label{fig:intrinsic}}

\end{document}


\newcommand\degd{\ifmmode^{\circ}\!\!\!.\,\else$^{\circ}\!\!\!.\,$\fi}
\newcommand{\etal}{{\it et al.\ }}
\newcommand{\uv}{(u,v)}
\newcommand{\rdm}{{\rm\ rad\ m^{-2}}}

\title{Detection of the Intrinsic Size of Sagittarius A* through 
Closure Amplitude Imaging:\\  
Supporting Online Material}

\author{Geoffrey C. Bower\altaffilmark{1}, 
Heino Falcke\altaffilmark{2},
Robeson M. Herrnstein\altaffilmark{3,4}
Jun-Hui Zhao\altaffilmark{3}, 
W.M. Goss\altaffilmark{5} \& 
Donald C. Backer\altaffilmark{1}}

\altaffiltext{1}{Astronomy Department \& Radio Astronomy Laboratory, 
University of California, Berkeley, CA 94720; gbower,dbacker@astro.berkeley.edu} 
\altaffiltext{2}{Radio Observatory Westerbork ,  ASTRON , P.O. Box 2 , 7990 AA Dwingeloo, The Netherlands; Visiting Scientist, Max-Planck-Institut fËr Radioastronomie, Auf dem H\"ugel 69, 53121 Bonn, Germany; Adjunct Professor, University of Nijmegen; falcke@astron.nl} 
\altaffiltext{3}{Department of Astronomy, Columbia University, Mail Code 5246, 500 West 120th St., New York, NY 10027; herrnstein@astro.columbia.edu}
\altaffiltext{4}{Harvard-Smithsonian Center for Astrophysics, 60 Garden Street, MS 78, Cambridge, MA 02138; jzhao@cfa.harvard.edu}
\altaffiltext{5}{National Radio Astronomy Observatory, Array Operations Center, P.O. Box O, Socorro, NM 87801; mgoss@aoc.nrao.edu}

We describe here further details of the analysis of this experiment.

Fringe fitting analysis in AIPS followed standard practices.
Single-band delay solutions obtained for NRAO 530 were transferred to Sgr A*.  
Multi-band delays and rates were determined by fringe-fitting Sgr A* itself. 
Fringes for Sgr A* were
found for stations Brewster (BR), Fort Davis (FD), Hancock (HN), Kitt Peak (KP), Los
Alamos (LA), North Liberty (NL), Owens Valley (OV), Pie Town (PT) and St. Croix (SC),
although not for all stations at all times. For instance, fringes with
BR are detected
at all wavelengths but only in the case of substantial fore-shortening at the 
longer wavelengths.  
While fringes were found between SC and HN, fringes were not found between either of 
these stations
and any other station forcing us  to exclude SC and HN from further analysis. 
We list all stations included in the analysis in
STable~\ref{tab:observations}.

We compute the closure amplitude for four stations $m, n, p$ and $q$:
\begin{equation}
C_{mnpq}= { |V_{mn}| |V_{pq}|  \over  |V_{mp}| |V_{nq}| },
\end{equation}
where $|V_{ij}|$ is the amplitude of the visibility on the baseline
between stations $i$ and $j$.  
The closure amplitude is closely related to the closure phase
for three stations $m, n$ and $p$:
\begin{equation}
\phi_{mnp}=\phi_{mn} + \phi_{np} + \phi_{pm}.
\end{equation}
Here $\phi_{ij}$ is the visibility phase on the baseline between stations $i$ and $j$.

The noise bias  principally has the effect of 
increasing model visibility amplitudes on long baselines where the Gaussian source 
is heavily
resolved.  
Our method of including the noise in the model visibilities avoids the problems of 
unbiasing the measured visibility amplitudes or the measured closure amplitudes.
The noise in the visibility amplitudes
biases the closure amplitudes by a factor that depends on
the unbiased visibility amplitude on that baseline  
\citep{1998ApJ...493..666T}.
The factor is accurate only for high SNR data and, therefore, not applicable
for our
case because our data is in the domain where SNR$\sim 1-10$.  A consequence
of our technique is that we are computing a quantity that differs slightly
from the true closure amplitude.

Time- and station-dependence of the noise bias is unimportant.  As the SNR approaches 1, the closure amplitudes have the largest error
and therefore the least weight in the $\chi^2$-fitting.
In the case that the noise bias is station- and time-independent, then the result is
dependent solely on the ratio $R=S_0/N$, where $S_0$ is the total flux density and $N$ is the noise bias
for a given integration time, typically 15 minutes.  This ratio ranged from $R\sim 10$
at 0.69 cm to $R\sim 100$ at 6.0 cm.  The flux density and the noise bias are not physical values 
since we have not performed any amplitude calibration.  
We found from examination of the data and Monte Carlo simulations 
that the ratio is determined with
about $5\%$ accuracy.  More importantly, the major axis, minor axis and position angle error
estimates are not strongly dependent on $R$; that is, 
these parameters are not covariant with $R$.
For example, the best-fit major axis value for the experiment BB130D is $708^{+17}_{-13} \ \mu$as.
At the $3\sigma$ extrema in $R$, the best-fit major axis values are $697$ and $704 \ \mu$as,
only marginally different.
This lack of covariance permits us to simplify our error methodology by 
dropping $R$ and considering only $x$, $y$ and $\phi$.

For the final results, we calculated $101^3$ grid points in
major axis, minor axis and position angle with a resolution of 5 $\mu$arcsec in major axis, 8
$\mu$arcsec in minor axis and $0.9$ degree in position angle at 0.69 cm.  Grid spacings in major
and minor axis increased with the square of wavelength for the other data sets.
Significance levels were determined assuming Gaussian statistics and computing
the appropriate increase in $\chi^2$ for the number of degrees of freedom
\citep{1969drea.book.....B}.

Our method for the determination of errors is not strictly correct since the error 
distributions for
the parameters are not normal.  This deviation
can be seen in the slices through the $\chi^2$ surface in SFigure~\ref{fig:ellipse}.  If
the distributions were normal, then the 3-dimensional error surface would be an ellipsoid.
Errors for each parameter are determined from the projection of the 3-dimensional surface 
into one dimension.
We report errors based on the 99.73\% confidence interval which corresponds to $3\sigma$ in
the normal case (STable~\ref{tab:results}).  When we cite errors throughout the rest of the paper,
we define the $1\sigma$ error as one-third of the $3\sigma$ error, not the $1\sigma$
error determined from the $\chi^2$ surface.  We do this because the $1\sigma$ error is not
as well determined from the $\chi^2$ surface.

We performed 
Monte Carlo simulations to relate $\Delta\chi^2$ to a confidence interval for the 0.69 cm experiment
BB130D.  We modeled errors in the data as Gaussian errors in the visibility amplitude after
vector averaging in wavelength and time.  We use the best-fit noise to determine the noise
distribution.  We also include station-dependent gain fluctuations as well as a time- and
station-dependent gain that represents opacity errors for a uniform atmosphere.  In the limit
that these errors do not change in closure amplitude averaging time (e.g., 15 minutes),
the closure amplitude method should be independent of these changes.  We used the best 
Gaussian parameters for the BB130D data (STable~\ref{tab:results}) 
as an input model for the visibilities and performed 1000 iterations
of the Monte Carlo test.  The resulting parameters fit to the data were distributed in
a Gaussian fashion and give means of $704 \pm 7 \ \mu$as in the 
major axis, $361 \pm 22 \ \mu$as in the minor axis and $81.4 \pm 1.3$ deg in the position
angle ($1\sigma$ errors).  The mean values are all consistent with the input model but  the errors
are lower by a factor of two to five from the errors determined by $\chi^2$ fitting.  
These Monte Carlo results set a firm lower limit on the errors.  
The apparent Gaussian nature of these errors and their small values relative to the errors determined
from the data, however, suggest that our Monte Carlo method does not fully
account for sources of error in the data.  Phase decorrelation effects not included
in the Monte Carlo simulation are the most likely source of the additional error.

All solutions discussed in the paper assume
$\beta=4$, as required by scattering theory.  We did, however, perform fits with $\beta$
unconstrained.  Not surprisingly, the other parameters were
determined with less accuracy when we added the additional parameter.
Nevertheless, their values were comparable with those previously
determined.  These individual measurements of $\beta$ had an error of
0.05 to 0.2.

We determine the mean value of $\beta$ 
as a function of wavelength (SFigure~\ref{fig:beta}).  The mean for all experiments is
$\beta=4.00 \pm 0.03$, fully consistent with the assumption of
Gaussian nature.  There is no evidence for a trend in $\beta$ with
wavelength.  We conclude that the images are precisely fit by a
Gaussian intensity distribution,  constraining the scattered image
to strictly follow a $\lambda^2$-law.  Any
deviation from a $\lambda^2$-law must indicate 
intrinsic structure.

Although the solutions for a single Gaussian component are sufficiently accurate, we 
did search the parameter space for two component models.  This search requires joint minimization
of closure phase and closure amplitude.  In all of the models considered,
the two Gaussians are assumed to be identical in major axis, minor axis and position
angle, which is the expectation of the
scattering model for the Galactic Center on these small angular scales
\citep{1994ApJ...427L..43F}.
For the most general case of this kind involving the relative position, where the three size
parameters and the flux densities  of the components are free parameters,
the solution from our minimization routine was highly sensitive to the initial guess.
Accordingly, we considered models in which the size and position
angle of the components are fixed at the L98 scattering value and the relative
position of the second component is fixed with respect to the main component
\citet{1998ApJ...508L..61L}.  At 0.69 cm,
the scattering model gives a size of 690 $\times$ 366 $\mu$as in a position angle of 80
degrees.  Only
the two flux density parameters and the noise bias parameter were variable for each model.  A $51 \times 51$
grid was constructed in relative position with steps equal to half the scattering size in
the given dimension.  In the case of 0.69 cm images, this corresponds to a field of 17.6 $\times$
8.8 mas.  This modeling gives the best-fit flux density of a second component
as a function of position.

In SFigure~\ref{fig:fluxrat}, we show four examples of flux density ratio maps.  These results can be
considered an upper limit on any second component.  For the case of BB130A, which is typical
of the 0.69 cm experiments, the flux density ratio peaks within the scattering size at $\sim 10\%$.  The
apparent double structure is spurious since the flux density ratio for the central point is
undefined.  The interpretation of this plot is that we cannot discriminate between the
single component Gaussian and two Gaussians separated by less than half the scattering
size with a flux density ratio of 10\% or less.  

We list in STable~\ref{tab:results} the maximum values of the flux
density ratio, $F_2$, in the maps, excluding the central few pixels
that are interior to the scattering region of the first source.  We
find that the maximum ratio at a given wavelength strongly scales with
wavelength.  The limits on secondary components are typically weaker
at long wavelengths, which is almost certainly a function of
the the more limited $\uv$-coverage available
for these experiments.  At 0.69 cm, limits on a secondary component are
typically 5\% while at 1.3 cm the limits are on the order of 10\%.

Due to the potentially significant effects of phase de-correlation,
we performed our complete analysis for averaging times of 15, 30, 60 and 120 seconds, 
providing a test of the dependence of the result on phase decorrelation.
We find sizes of $22 \pm 1$, $23 \pm 2$, $24 \pm 2$ and $24\pm 2$  $R_s$ 
for the intrinsic size of Sgr A* at 0.69 cm
for each of these averaging times, respectively.  Thus, the results are not biased by
our averaging time at a level more than $1 R_s$.

We also show that no one experiment determines the significance of the results.
Dropping experiment BB130F, which has the smallest errors, we find that the mean
0.69 cm size is $716^{+17}_{-12}\ \mu$sec.  This size is only slightly larger than
the mean size of $712^{+12}_{-9}\ \mu$sec.  The resulting discrepancy with the
L98 scattering law is $6\sigma$ instead of $7\sigma$.  Dropping both BB130F and
BB130B from the average, we find a size $714^{+17}_{-12}\ \mu$sec, which is
also strongly inconsistent with the L98 scattering law.

We summarize in STable~\ref{tab:scattering} the results of fitting the major and minor axis sizes as a function of wavelength.  We solve for the power-law index, $\alpha$, and the normalized size at a wavelength of 1 cm, $\sigma^{\rm 1cm}$, for each axis as a function of the minimum wavelength, $\lambda_{min}$, included in the fit.  Fits are included with $\alpha$ constrained to 2 and unconstrained.  We also compute the reduced $\chi^2_{\nu}$ for each fit.
  
\bibliographystyle{apj}
\bibliography{myrefs}

\plotone{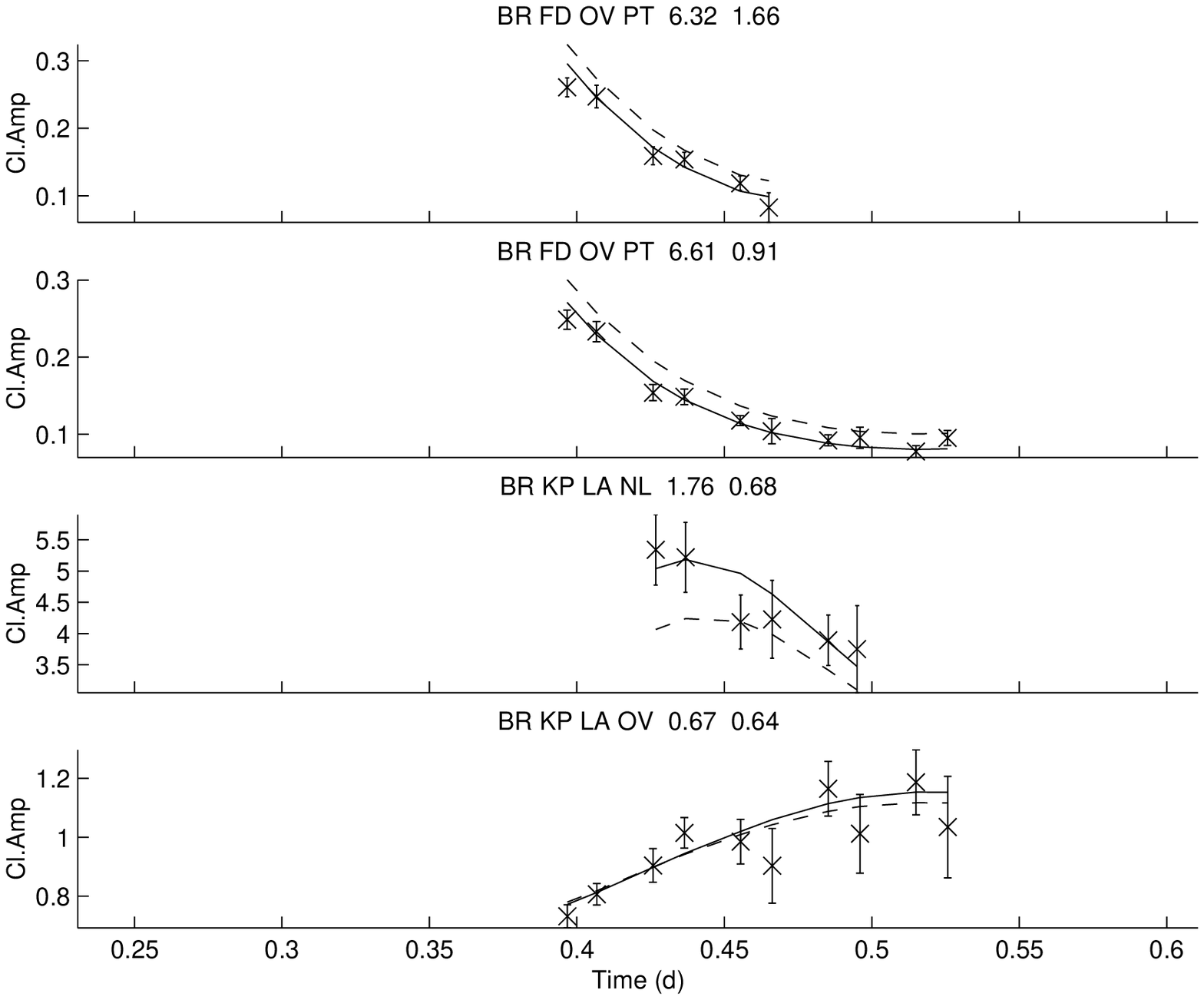}
\figcaption{Closure amplitude for four station groups from the 0.69 cm experiment BB130D.
The numbers after the station identification are the
reduced $\chi^2$ for the  L98 scattering model (dashed line) and the best fit model
(solid line).  The improvement in the model fit is readily apparent for these baselines.
The station group BR-FD-OV-PT appears twice because  two independent
closure amplitudes can be formed from the set of four stations.  These have different
extent in time because some baselines exceed the 250$M\lambda$ cutoff for baseline 
length.
\label{fig:clamp}
}

\plotone{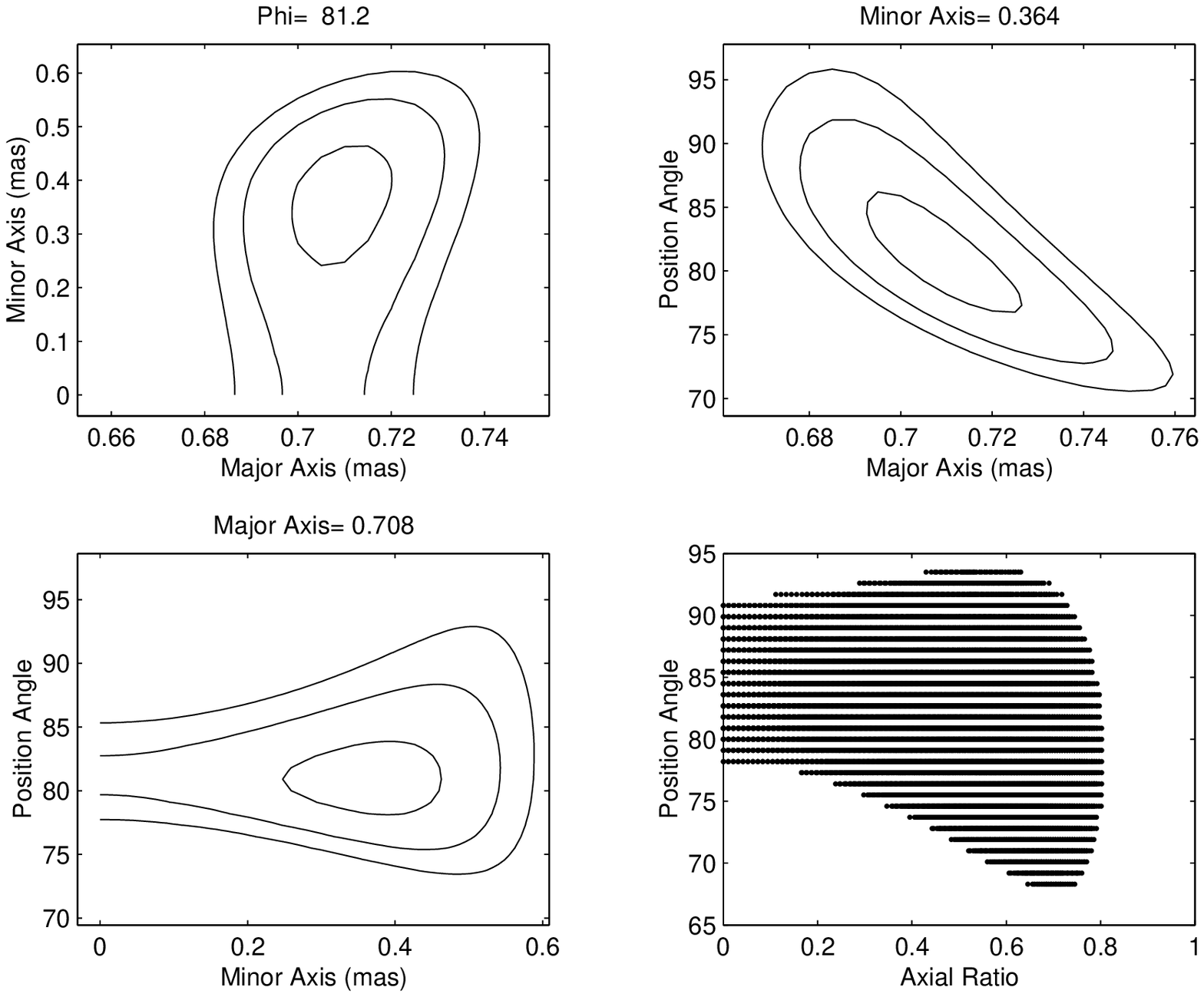}
\figcaption{Three cuts through the $\chi^2$ surface in major axis, minor axis and position angle space
for 68\%, 95.4\%, and 99.73\% confidence intervals
of source parameters for one 0.69 cm observation (BB130D) analysed with the closure amplitude method.  
These contours correspond to 1, 2 and 3$\sigma$ for the case of Gaussian errors.  Cuts are 
centered on the best-fit values of each parameter, which are listed above the plot.  
The lower right corner shows
grid points of 99.73\% confidence interval in axial ratio versus position angle space.
\label{fig:ellipse}
}

\plotone{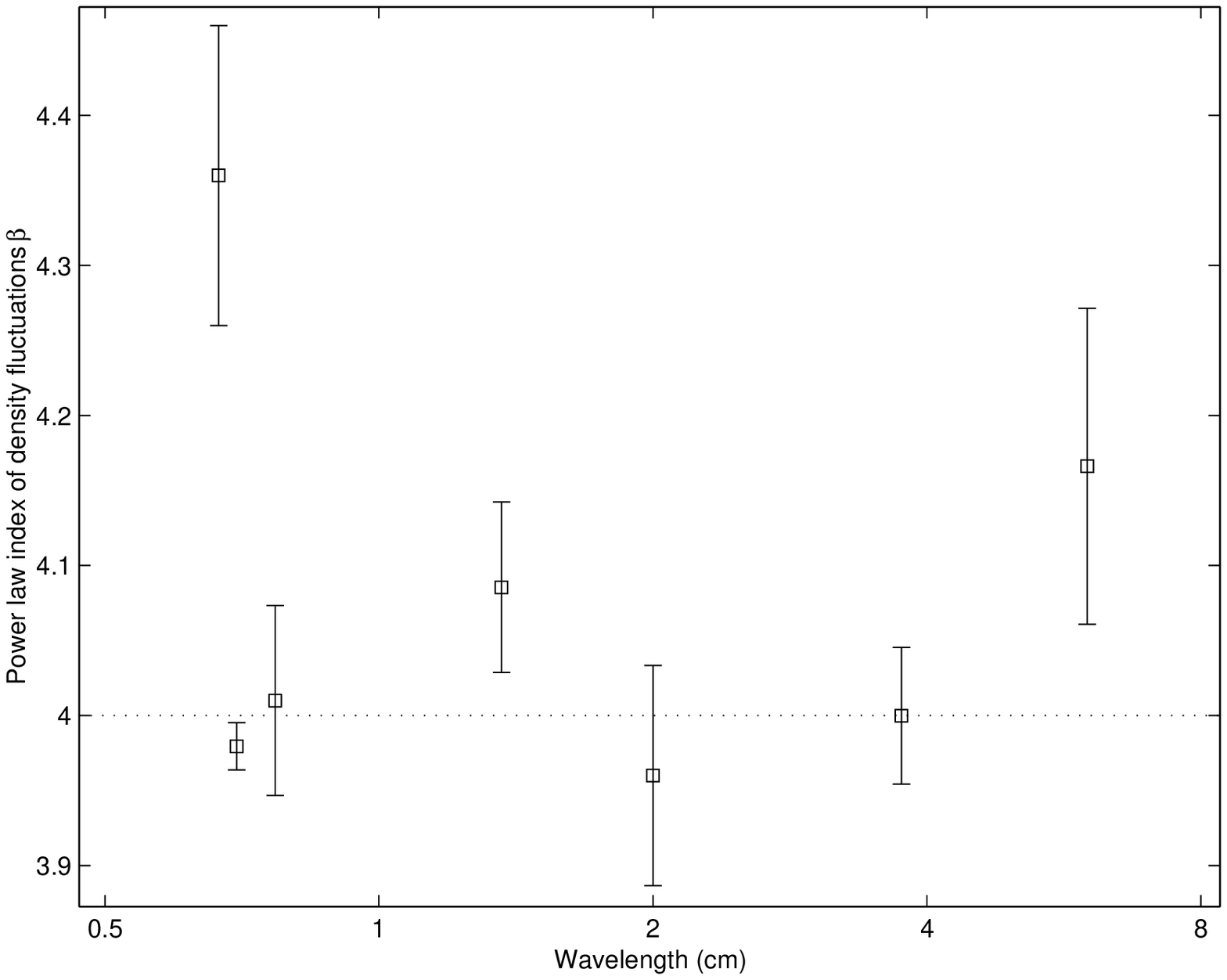}
\figcaption{The power-law index of density fluctuations, $\beta$, plotted as a function
of wavelength.  Individual results and the mean value over all wavelengths $< \beta > = 4.00 \pm 0.03$ are consistent with strong scattering on baselines shorter than the inner scale of turbulence.
\label{fig:beta}
}

\plotone{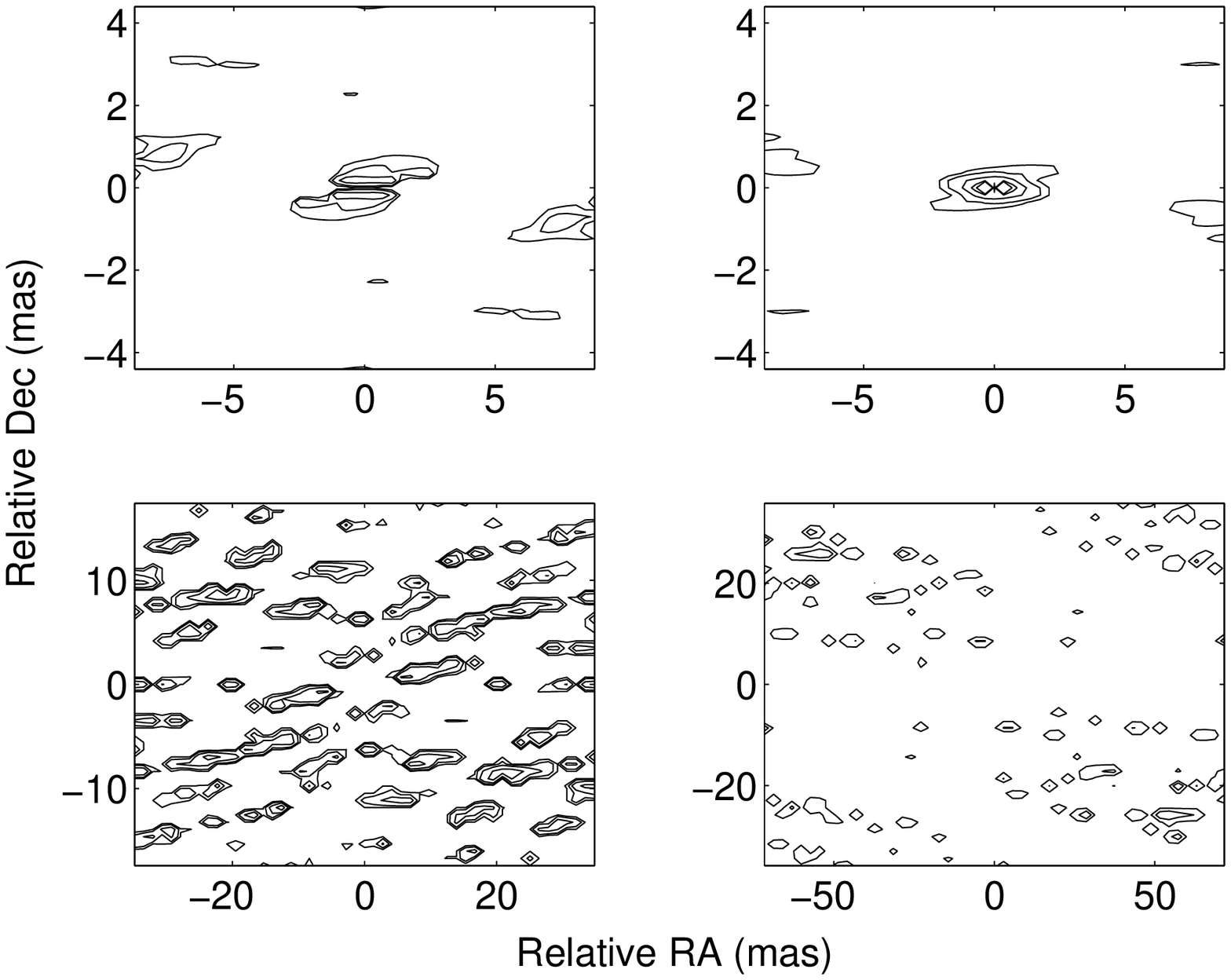}
\figcaption{Maps of the best-fit flux density ratio for a two Gaussian component model as a function of
relative position.  Contour intervals are 1, 2, 4, 8 and 16\%.  The maps are for the 0.69 cm
experiments BB130D (upper left) and BB130F (upper right), 1.3 cm experiment BB130A (lower left)
and 2.0 cm experiment BS055C (lower right). 
\label{fig:fluxrat}
}

\begin{deluxetable}{lrrccccccccccc}
\tablecaption{Observations of Sgr A* \label{tab:observations}}
\tablehead{\colhead{Code} & \colhead{Date} & \colhead{Freq.} & 
\colhead{BR} & \colhead{FD} &  \colhead{KP} & \colhead{LA} & \colhead{NL} & \colhead{OV} & \colhead{PT} &  \colhead{Y1} \\
                          & \colhead{(dd/mm/yy)} & \colhead{(GHz)}  }
\startdata
\hline
\multicolumn{4}{c}{New Observations} \\
\hline
BB130A & 12/07/01 & 43 & $\surd$ & $\surd$ & $\surd$ & $\surd$ & $\surd$ & $\surd$ & $\surd$ & \dots \\ 
\dots  & \dots    & 22 & $\surd$ & $\surd$ & $\surd$ & $\surd$ & \dots & $\surd$ & $\surd$ & \dots \\ 
BB130B & 29/07/01 & 43 & $\surd$ & $\surd$ & $\surd$ & $\surd$ & $\surd$ & $\surd$ & $\surd$ & \dots \\ 
\dots  & \dots    & 22 & $\surd$ & $\surd$ & $\surd$ & $\surd$ & $\surd$ & $\surd$ & $\surd$ & \dots \\ 
BB130C & 05/08/01 & 43 &$\surd$ & $\surd$ & $\surd$ & $\surd$ & $\surd$ & $\surd$ & $\surd$ & \dots \\ 
\dots  & \dots    & 22 & $\surd$ & $\surd$ & $\surd$ & $\surd$ & $\surd$ & $\surd$ & $\surd$ & \dots \\ 
BB130D & 15/04/02 & 43 & $\surd$ & $\surd$ & $\surd$ & $\surd$ & $\surd$ & $\surd$ & \dots & \dots \\
BB130F & 03/05/02 & 43 &$\surd$ & $\surd$ & $\surd$ & $\surd$ & $\surd$ & $\surd$ & $\surd$ & \dots \\ 
BB130G & 13/05/02 & 43 & $\surd$ & $\surd$ & $\surd$ & $\surd$ & $\surd$ & $\surd$ & $\surd$ & \dots \\ 
\hline
\multicolumn{4}{c}{Archival Data} \\
\hline
BS055A & 07/02/97 & 4.98  &  \dots   & $\surd$ & $\surd$ & $\surd$ & \dots & \dots & $\surd$ & $\surd$ \\ 
\dots  & \dots    & 8.42  &  \dots   & $\surd$ & $\surd$ & $\surd$ & \dots   & \dots & $\surd$ & $\surd$ \\ 
BS055B & 12/02/97 & 22  &  \dots   & $\surd$ & $\surd$ & $\surd$ & $\surd$ & $\surd$ & $\surd$ & $\surd$ \\ 
\dots  & \dots    & 15  &  \dots   & $\surd$ & $\surd$ & $\surd$ & \dots   & $\surd$ & $\surd$ & $\surd$ \\ 
BS055C & 14/02/97 & 43  &  $\surd$ & $\surd$ & $\surd$ & $\surd$ & $\surd$ & $\surd$ & \dots   & $\surd$ \\ 
BL070B & 23/05/99 & 43 &  $\surd$ & $\surd$ & $\surd$ & $\surd$ & $\surd$ & $\surd$ & \dots   & \dots   \\ 
BL070C & 31/05/99 & 45 &  $\surd$ & $\surd$ & $\surd$ & $\surd$ & $\surd$ & $\surd$ & \dots   & \dots   \\ 
\dots  & \dots    & 39 &  $\surd$ & $\surd$ & $\surd$ & $\surd$ & $\surd$ & $\surd$ & \dots   & \dots   \\ 
BB113  & 29/08/99 & 8.42 & $\surd$ & $\surd$ & $\surd$ & $\surd$ & $\surd$ & $\surd$ & $\surd$ & $\surd$ \\ 
\dots & \dots     & 4.99 & \dots   & $\surd$ & $\surd$ & $\surd$ & $\surd$ & $\surd$ & $\surd$ & $\surd$ \\ 
\enddata
\end{deluxetable}

\begin{deluxetable}{llrrrrrr}
\tablecaption{Closure Amplitude and Closure Phase Modeling Results\label{tab:results}}
\tablehead{\colhead{Code} & \colhead{$\nu$} & \colhead{$x$} & 
\colhead{$y$} & 
\colhead{$\phi$} & 
\colhead{$\chi^2_{A}$}  &
\colhead{$\chi^2_{\phi}$} &
\colhead{$F_{2}$} \\
                          & \colhead{(GHz)} & \colhead{($\mu$as)} & \colhead{($\mu$as)} & \colhead{(deg)} } 
\startdata
\hline
\multicolumn{7}{c}{5.0 GHz} \\
\hline
BS055A   &   5.0 & $48638^{+6562}_{-4838}$ & $15381^{+17619}_{-15381}$ & $ 87.1^{+10.9}_{-15.2}$ &  2.0  & 1.7 & 0.31\\
   BB113 &  5.0 & $ 47850^{+10950}_{-5250} $ & $ 24950^{+17650}_{-9950} $ & $  73^{+ 28}_{- 15} $ &  2.2 & 1.0 & 0.02\\
ALL      &   5.0 & $48359^{+5628}_{-3559}$ & $21189^{+12468}_{-\ 8356}$ & $ 83.2^{+10.2}_{-10.6}$ &  \dots  & \dots & \dots\\
\hline
\multicolumn{7}{c}{8.4 GHz} \\
\hline
BS055A &   8.4 & $17697^{+ 303}_{- 297}$ & $7804^{+1596}_{-2204}$ & $ 81.8^{+ 4.5}_{- 4.5}$ &  0.9 & 3.3 &  0.03\\
    BB113 &  8.4 & $ 16840^{+960}_{-1040} $ & $ 10150^{+2050}_{-3950} $ & $  83^{+ 13}_{- 18} $ &  2.1 & 1.3 &   0.17 \\
ALL    &   8.4 & $17626^{+ 289}_{- 286}$ & $8514^{+1260}_{-1925}$ & $ 81.9^{+ 4.3}_{- 4.4}$ &  \dots & \dots & \dots\\

\hline
\multicolumn{7}{c}{15 GHz} \\
\hline
BS055B &  15.3 & $5198^{+ 262}_{- 228}$ & $1963^{+2027}_{-1963}$ & $ 84.0^{+11.3}_{-11.2}$ &  1.4 & 1.1 & 0.03\\
\hline
\multicolumn{7}{c}{22 GHz} \\
\hline
BB130A &  22.2 & $2817^{+ 683}_{- 317}$ & $1732^{+ 468}_{-1732}$ & $ 70.9^{+21.7}_{-23.3}$ &  2.7 & 1.3  & 0.11\\
BB130B &  22.2 & $2655^{+ 345}_{- 155}$ & $1736^{+ 464}_{- 922}$ & $ 74.2^{+13.9}_{-21.2}$ &  2.0 & 0.9  & 0.09\\
BB130C &  22.2 & $2852^{+ 518}_{- 312}$ & $1721^{+ 435}_{- 929}$ & $ 63.0^{+17.0}_{-18.1}$ &  2.2  & 1.2  & 0.15 \\
BS055B &  22.2 & $2607^{+ 123}_{-\  87}$ & $1164^{+ 552}_{-1164}$ & $ 77.8^{+\ 6.7}_{-10.4}$ &  0.6 & 1.8 & 0.01\\
ALL      & 22.2 & $2635^{+ 112}_{-\ 72}$ & $1613^{+ 237}_{- 542}$ & $74.6^{+\ 5.6}_{-\ 7.8}$ & \dots & \dots  & \dots\\
\hline
\multicolumn{7}{c}{39 GHz} \\
\hline
BL070C &  39.1 & $ 884^{+\  16}_{-\  44}$ & $ 610^{+ 140}_{- 340}$ & $ 70.4^{+13.2}_{-20.1}$ &  1.7 & 1.0  & 0.03\\
\hline
\multicolumn{7}{c}{43 GHz} \\
\hline
BB130A &  43.2 & $ 725^{+\  65}_{-\  35}$ & $ 498^{+ 139}_{- 296}$ & $ 75.7^{+13.3}_{-26.3}$ &  1.3 & 1.1  & 0.03\\
BB130B &  43.2 & $ 770^{+\  90}_{-\  55}$ & $ 465^{+ 180}_{- 465}$ & $ 69.6^{+14.0}_{-26.5}$ &  1.0 & 1.1  & 0.06\\
BB130C &  43.2 & $ 704^{+ 191}_{- 129}$ & $ 323^{+ 427}_{- 323}$ & $ 76.2^{+48.8}_{-41.2}$ &  1.7 & 0.9  & 0.07\\
BB130D &  43.2 & $ 708^{+\  52}_{-\  38}$ & $ 364^{+ 221}_{- 364}$ & $ 81.2^{+12.3}_{-12.9}$ &  0.8 & 1.3 & 0.04\\
BB130F &  43.2 & $ 708^{+\  17}_{-\  13}$ & $ 360^{+ 120}_{- 218}$ & $ 81.4^{+\ 4.9}_{-\ 6.8}$ &  0.9 & 1.2 & 0.03\\
BB130G &  43.2 & $ 709^{+\  26}_{-\  19}$ & $ 350^{+ 153}_{- 350}$ & $ 81.4^{+\ 5.8}_{-\ 9.5}$ &  0.7 & 1.3  & 0.03\\
BL070B &  43.1 & $ 713^{+\  37}_{-\  28}$ & $ 434^{+ 129}_{- 246}$ & $ 76.6^{+\ 8.8}_{-15.5}$ &  0.9 & 1.0  & 0.02\\
BS055C &  43.2 & $ 728^{+\  47}_{-\  33}$ & $ 428^{+ 157}_{- 428}$ & $ 75.8^{+\ 9.6}_{-17.4}$ &  0.5 & 1.9  & 0.03\\
ALL & 43.2 & $712^{+\ 12}_{-\ \ 9}$ & $407^{+\  56}_{- 109}$ & $79.8^{+\ 3.0}_{-\ 4.5}$ & \dots & \dots & \dots\\
\hline
\multicolumn{7}{c}{45 GHz} \\
\hline
BL070C5 &  45.1 & $ 683^{+ 132}_{-  73}$ & $ 458^{+ 292}_{- 458}$ & $ 76.5^{+8.5}_{-41.5}$ &  1.2  & 1.3 & 0.03\\
\enddata
\tablecomments{$x$ is the major axis size of the single elliptical Gaussian component;
$y$ is the minor axis size; $\phi$ is the position angle; $\chi^2_A$ is the reduced
$\chi^2$ for the closure amplitudes for a single component elliptical Gaussian; $\chi^2_\phi$
is the reduced $\chi^2$ for the closure phases for an axisymmetric model; and, $F_2$ is the 
maximum ratio of the flux densities in a two component model. {\em Errors are $3\sigma$. }} 
\end{deluxetable}

\begin{deluxetable}{lrrrrrrrr}
\tablecaption{Fits to the Size of Sgr A* as a Function of Wavelength \label{tab:scattering}}
\tablehead{
\colhead{$\lambda_{\rm min}$} &
\colhead{$\alpha_{\rm major}$} & \colhead{$\sigma_{\rm major}^{\rm 1cm}$} & \colhead{$\chi^2_\nu$} & \colhead{d.o.f.} &
\colhead{$\alpha_{\rm minor}$} & \colhead{$\sigma_{\rm minor}^{\rm 1cm}$} & \colhead{$\chi^2_\nu$} & \colhead{d.o.f.} \\
\colhead{(cm)} &                   & \colhead{ (mas) }        &                    &                  &
                   & \colhead{ (mas) }
}
\startdata
2.0 & $ 2.01^{0.03}_{0.03} $ & $ 1.37^{0.05}_{0.04} $ &   1.6  &  1 &$ 2.01^{0.15}_{0.22} $ & $ 0.64^{0.25}_{0.08} $ &   1.1 & 1 \\
\dots & $ 2  $ & $ 1.39^{0.01}_{0.01} $ &   1.0  &  2 & 2  & $ 0.65^{0.04}_{0.05} $ &   0.5 & 2 \\
1.3 & $ 1.96^{0.01}_{0.01} $ & $ 1.45^{0.02}_{0.02} $ &   2.9  &  2 &$ 1.75^{0.15}_{0.09} $ & $ 0.91^{0.08}_{0.13} $ &   1.1 & 2 \\
\dots & $ 2  $ & $ 1.40^{0.01}_{0.01} $ &   5.6  &  3 & 2  & $ 0.68^{0.03}_{0.04} $ &   2.0 & 3 \\
0.6 & $ 1.96^{0.01}_{0.00} $ & $ 1.46^{0.01}_{0.00} $ &   2.5  &  5 &$ 1.85^{0.06}_{0.06} $ & $ 0.81^{0.03}_{0.05} $ &   0.9 & 5 \\
\dots & $ 2  $ & $ 1.44^{0.01}_{0.00} $ &  18.3  &  6 & 2  & $ 0.71^{0.03}_{0.04} $ &   2.3 & 6 \\
0.3 & $ 1.96^{0.01}_{0.00} $ & $ 1.46^{0.01}_{0.01} $ &   2.1  &  6 &$ \dots $ & \dots &   \dots & \dots \\
\dots & $ 2  $ & $ 1.44^{0.01}_{0.01} $ &  15.7  &  7 & \dots  & \dots &   \dots  & \dots \\
\enddata
\end{deluxetable}